\documentclass[a4paper,fleqn,usenatbib]{mnras}

\usepackage[T1]{fontenc}
\usepackage{ae,aecompl}

\usepackage{graphicx}   
\usepackage{amsmath}    
\usepackage{amssymb}    
\usepackage{float}
\usepackage{subfig}

\newcommand{\Mpc}{\mathrm{~km~s^{-1}~Mpc^{-1}}}

\def\({\left(}
\def\){\right)}
\def\[{\left[}
\def\]{\right]}

\title[Probing modified gravity with high-redshift quasars]{Probing modified gravity theories with multiple measurements of high-redshift quasars}
\author[Lian et al.]
{Yujie Lian$^{1}$, Shuo Cao$^{1}$,\thanks{E-mail:
caoshuo@bnu.edu.cn} Marek Biesiada$^{2}$,\thanks{E-mail:
Marek.Biesiada@ncbj.gov.pl} Yun Chen$^{3}$, Yilong Zhang$^{1}$, Wuzheng Guo$^{1}$ \\
$^{1}$ Department of Astronomy, Beijing Normal University, 100875 Beijing, China; \\
$^{2}$ National Centre for Nuclear Research, Pasteura 7, 02-093 Warsaw, Poland;\\
$^3$ National Astronomical Observatories, Chinese Academy of
Sciences, 100012 Beijing, China}

\date{Accepted XXX. Received YYY; in original form ZZZ}

\pubyear{2021}

\begin{document}
\label{firstpage}
\pagerange{\pageref{firstpage}--\pageref{lastpage}}
\maketitle

\begin{abstract}
In this paper, we quantify the ability of multiple measurements of
high-redshift quasars (QSOs) to constrain several theories of
modified gravity, including the Dvali-Gabadadze-Porrati braneworld
scenario, generalized Chaplygin gas, $f(T)$ modified gravity, and
Modified Polytropic Cardassian model. Recently released sample of
1598 quasars with X-ray and UV flux measurements in the redshift
range of $0.036\leqslant z \leqslant 5.1003$, as well as a
compilation of 120 intermediate-luminosity radio quasars covering
the redshift of $0.46<z<2.76$ are respectively used as standard
probes at higher redshifts. For all considered modified gravity
theories, our results show that there is still some possibility that
the standard $\Lambda$CDM scenario might not be the best
cosmological model preferred by the current quasar observations. In
order to improve cosmological constraints, the quasar data are also
combined with the latest observations of baryon acoustic
oscillations (BAO), which strongly complement the constraints.
Finally, we discuss the support given by the data to modified
gravity theories, applying different information theoretic
techniques like the Akaike Information Criterion (AIC), Bayesian
Information Criterion (BIC) and Jensen-Shannon divergence (JSD).
\end{abstract}


\section{Introduction}

The discovery of the accelerating expansion of the universe, first
confirmed by observations of type Ia supernovae (SN Ia)
\citep{Riess1998,Perlmutter1999}, is a milestone in modern cosmology
and has since been verified by other cosmological observations,
including the cosmic microwave background (CMB) \citep{Spergel2003},
BAO \citep{Eisenstein2005,Percival2007}, large-scale structure
\citep{Tegmark2004}. However, there are different understandings
about the origin of cosmic acceleration, which has led to many
cosmological scenarios principally based on two large categories
being proposed and developed. On the one hand, in the framework of
Einstein's Theory of General Relativity a mysterious component with
negative pressure, dubbed dark energy (DE) \citep{Copeland2006},
responsible for the accelerated cosmological expansion is proposed.
On the other hand, modifying the theory of gravity
\citep{Tsujikawa2010} is another direction to understand this
phenomenon instead of adding new hypothetical material components.

In the first scenario, the simplest candidate for dark energy is the
cosmological constant $\Lambda$, a modification of the
energy-momentum tensor in Einstein equations, which is constant in
time and underlies the simplest standard cosmological model -- the
$\Lambda$CDM model. While $\Lambda$CDM is consistent with many
observations
\citep{Allen2008,Cao2012,Alam2017,Farooq2017,Scolnic2018}, this
model is still confronted with some theoretical problems such as the
well-known fine-tuning problem and coincidence problem
\citep{Weinberg1989}, which has prompted a great number of dark
energy models including dynamic dark energy models
\citep{Boisseau2000,Kamenshchik2001,Maor2001}, interacting dark
energy model \citep{Amendola2000,Caldera-Cabral2009} and scalar
field theories
\citep{Peebles1988,Ratra1988,Zlatev1999,Caldwell2005,Chen2011,Chen2016},
to be proposed and studied. In the second scenario, many modified
gravity theories not only provides interesting ideas to deal with
the cosmological constant problem and explain the late-time
acceleration of the universe without DE but also describe the large
scale structure distribution of the universe (see
\citet{Clifton2012,Koyama2016} for recent reviews). One idea to
modify gravity is assuming that our universe is embedded in a higher
dimensional spacetime, such as the brane-world
Dvali-Gabadadze-Porrati (DGP) model \citep{Dvali2000,Sollerman2009},
modified polytropic Cardassian (MPC) model
\citep{Wang2003,Magana2015}, and Gauss-Bonnet gravity
\citep{Nojiri2005}. Another interesting idea is to extend General
Relativity (GR) by permitting the field equation to be higher than
second order, like $f(R)$ gravity \citep{Chiba2003,Sotiriou2010}, or
change the Levi-Civita connection to the Weitzenb\"ock connection
with torsion, such as $f(T)$ gravity
\citep{Bengochea2009,Yang2011,Cai2016}. In this paper, we
concentrate on four cosmological models in the framework work of
Friedman-Lema\^{\i}tre-Robertson-Walker metric, including
Generalized Chaplygin Gas (GCG) model, a kind of dynamical dark
energy model, in which the dark energy density decreases with time,
DGP model, MPC model, and the power-law $f(T)$ model, based on
teleparallel gravity.

With so many competitive cosmological models, many authors have
taken advantage of various cosmological probes, such as SN Ia
\citep{Nesseris2005,Suzuki2012,Scolnic2018}, Gamma-ray burst
\citep{Lamb2000,Liang2005,Ghirlanda2006,Rezaei2020}, HII starburst
galaxies
\citep{Siegel2005,Plionis2011,Terlevich2015,Wei2016,Wu2020,Cao2020}
acting as standard candles, strong gravitational lensing systems
\citep{Biesiada2011,Cao11,Cao12a,Cao12b,Cao2012,Cao2015,Chen2015,Cao17c,Liu19,Amante2020},
galaxy clusters \citep{Bonamente2006,De Bernardis2006,Chen2012}, BAO
measurements, CMB \citep{Spergel2003,Planck2016,Planck2018} acting
as standard rulers to test these models or in other similar
cosmological studies. Furthermore, it is crucial to test which model
is most favored by current observations, in addition to the most
important aim that is to constrain cosmological parameters more
precisely. To fulfill this tough goal, better and diverse data sets
are required.

Recently, quasars observed with multiple measurements, another
potential cosmological probe with a higher redshift range that
reaches to $z \sim 5$, is becoming popular to constrain cosmological
models in the largely unexplored portion of redshift range from  $z
\sim 2$ to $z \sim 5$. A sample that contains 120 angular size
measurements in intermediate-luminosity quasars from the very-long
baseline interferometry (VLBI) observations \citep{Cao17a,Cao17b},
has become an effective standard ruler, which have been extensively
applied to test cosmological models
\citep{Qi2017,Melia17,Li17,Zheng2017,Xu2018,Ryan2019}, measuring the
speed of light \citep{Cao17a,Cao20} and exploring cosmic curvature
at different redshifts \citep{Qi2019,Cao19} and the validity of
cosmic distance duality relation \citep{Zheng2020}. Then,
\citet{Risaliti2019} put forward a new compilation of quasars
containing 1598 QSO X-Ray and UV flux measurements in the redshift
range of $0.036  \leqslant z  \leqslant 5.1003$, which have been
used to constrain cosmological models \citep{Khadka2020b} and cosmic
curvature at high redshifts \citep{LiuTH20a,LiuYT2020}, as well as
test the cosmic opacity \citep{LiuTH20b,Geng2020}. Making use of
this data to explore cosmological researches mainly depends on the
empirical relationship between the X-Ray and UV luminosity of these
high redshift quasars proposed by \citet{Avni1986}, which leads to
the Hubble diagram constructed by quasars
\citep{Risaliti2015,Lusso2016,Risaliti2017,Bisogni2018}. In general,
the advantage of these two QSO measurements over other traditional
cosmological probes is that QSO has a larger redshift range, which
may be rewarding in exploring the behavior of the non-standard
cosmological models at high redshifts, providing an important
supplement to other astrophysical observations and also
demonstrating the ability of QSO as an additional cosmological probe
\citep{Zheng21}.

In this paper, we focus on applying the angular size measurements of
intermediate-luminosity quasars \citep{Cao17a,Cao17b} and the large
QSO X-ray and UV flux measurements \citep{Risaliti2019} to constrain
four non-standard cosmological models, with the main goal of testing
the agreement between the high-redshift combined QSO data and the
standard $\Lambda$CDM model through the performance of these
non-standard models at higher redshift, as well as demonstrating the
potential of QSO as an additional cosmological probe. In order to
make the constraints more stringent and test consistency, 11 recent
BAO measurements \citep{Cao2020} are considered in the joint
analysis with the combined QSO measurements, at the redshift range
$0.122  \leq z \leq 2.34$. This paper is organized as follows. In
Sec. 2, all the observations we used in this work are briefly
introduced. In Sec. 3, we describe the non-standard cosmological
models we considered, and details of the methods used to constrain
the model parameters are described in Sec. 4. In Sec.5, we perform a
Markov chain Monte Carlo (MCMC) analysis using different data sets,
and apply some techniques of model selection. Finally, conclusions
are summarized in Sec. 6.

\begin{table}
        \centering
        \caption{The BAO data. Distances $D_{M}(r_{s,fid}/r_{s})$, $D_{V}(r_{s,fid}/r_{s})$, $r_{s}$ and $r_{s,fid}$ have the units of $Mpc$, while $H(z)(r_{s}/r_{s,fid})$ has the units of $km\,s^{-1}\,Mpc^{-1}$, and $D_{A}/r_{s}$, $D_{H}/r_{s}$, as well as $D_{M}/r_{s}$ are dimensionless. \textbf{The correlation matrix of the six measurements from \citet{Alam2017} and the two measurements from \citet{de2019} can be found in \citet{Ryan2019} and \citet{Cao2020} respectively.}}
        \label{tab:BAOdata_table}
        \setlength{\tabcolsep}{1mm}{
            \begin{tabular}{lccr} 
                \hline
                z& Measurement & Value & Ref.\\
                \hline
                0.38 & $D_{M}(r_{s,fid}/r_{s})$ & 1512.39 & \citet{Alam2017}\\
                \hline
                0.38 & $H(z)(r_{s}/r_{s,fid})$ & 81.2087 & \citet{Alam2017}\\
                \hline
                0.51 & $D_{M}(r_{s,fid}/r_{s})$ & 1975.22 & \citet{Alam2017}\\
                \hline
                0.51 & $H(z)(r_{s}/r_{s,fid})$ & 90.9029 & \citet{Alam2017}\\
                \hline
                0.61 & $D_{M}(r_{s,fid}/r_{s})$ & 2306.68 & \citet{Alam2017}\\
                \hline
                0.61 & $H(z)(r_{s}/r_{s,fid})$ & 98.9647 & \citet{Alam2017}\\
                \hline
                0.122 & $D_{V}(r_{s,fid}/r_{s})$ & $539 \pm 17$ & \citet{Carter2018}\\
                \hline
                0.81 & $D_{A}/r_{s}$ & $10.75 \pm 0.43$ & \citet{DES2019}\\
                \hline
                1.52 & $D_{V}(r_{s,fid}/r_{s})$ & $3843 \pm 147$ & \citet{Ata2018}\\
                \hline
                2.34 & $D_{H}/r_{s}$ & 8.86 & \citet{de2019}\\
                \hline
                2.34 & $D_{M}/r_{s}$ & 37.41 & \citet{de2019}\\
                \hline
        \end{tabular}}
    \end{table}

    \section{DATA}

Quasars are one of the brightest sources in the universe. Observable
at very high redshifts, they are regarded as particularly promising
cosmological probes. In the past decades, different relations
involving the quasar luminosity have been proposed to study the
"redshift - luminosity distance" relation in quasars with the aim of
cosmological applications \citep{Baldwin1977,Watson2011,Wang2013}.
Accordingly, \citet{Risaliti2015} compiled a sample of 808 quasar
flux-redshift measurements over a redshift range $0.061  \leq z \leq
6.280$ with the aim to constrain cosmological models. More than
three-quarters of quasars in this sample are located at high
redshift ($z > 1$). It is worth to notice that this compilation
alone did not give very tight constraints on the cosmological
parameters compared with other data \citep{Khadka2020a}, on account
of the large global intrinsic dispersion ($\delta = 0.32$) in the
X-ray and UV luminosity relation. Recently, \citet{Risaliti2019}
proposed a final compilation of 1598 quasars flux-redshift
measurements, selected from a sample of 7238 quasars with available
X-ray and UV measurements, to find more high-quality quasars
applicable to cosmological research. Compared with the 2015 data
set, the latest quasar sample has a smaller redshift range ($0.036
\leq z \leq 5.1003$) whereas 899 quasars at high redshift ($z > 1$)
are included in the sample. Meanwhile, with the progressively
refined selection technique, flux measurements, and the efforts of
eliminating systematic errors, the Hubble diagram produced by this
large quasar sample is in great accordance with that of supernovae
and the concordance model at $z \leq 1.4$ \citep{Risaliti2019}.
Besides, these QSOs have an X-ray and UV luminosity relation with a
smaller intrinsic dispersion ($\delta = 0.23$).

Besides the X-ray and UV flux measurements of quasars, we also use
the angular size measurements in radio quasars
\citep{Cao17a,Cao17b,Cao18}, from the very-long-baseline
interferometry (VLBI) observations, which has become a reliable
standard ruler in cosmology. The measurements of
milliarcsecond-scale angular size from compact radio sources
\citep{Gurvits1999} have been utilized for cosmological models
inference \citep{Vishwakarma2001,Zhu2002,Chen2003}. Notably, the
angular size measurements are effective only if the linear size
$l_m$ of the compact radio sources is independent on both redshifts
and intrinsic properties of the source such as luminosity. More
recently, \citet{Cao17b} presented a final sample of 120
intermediate-luminosity quasars ($10^{27} W/Hz \leq L \leq 10^{28}
W/Hz$) over the redshift range $0.46  < z < 2.76$ from VLBI all-sky
survey of 613 milliarcsecond ultra-compact radio sources
\citep{Kellermann1993,Gurvits1994}, in which these
intermediate-luminosity quasars show negligible dependence on
redshifts and intrinsic luminosity. Meanwhile, a
cosmology-independent method to calibrate the linear size as $l_m =
11.03 pc$ was implemented in the study and these angular size versus
redshift data have been used to constrain cosmological parameters.

Additionally, in order to acquire smaller uncertainty, as well as to
compare the constraints to the joint analysis with other
cosmological probes, we also add 11 recent BAO data \citep{Cao2020}
in our analysis. These data come from the large scale structure
power spectrum through astronomical surveys and have been
extensively applied in cosmological applications, covering the
redshift range $0.122  \leq z \leq 2.34$. Figure 1 indicates the
redshift distribution of the QSO measurements and BAO data, where we
display the X-ray and UV fluxes QSO measurements, the angular size
measurements in radio quasars and BAO measurements by the
abbreviation QSO[XUV], QSO[AS], and BAO respectively.

\begin{figure}
\includegraphics[width=0.52\textwidth]{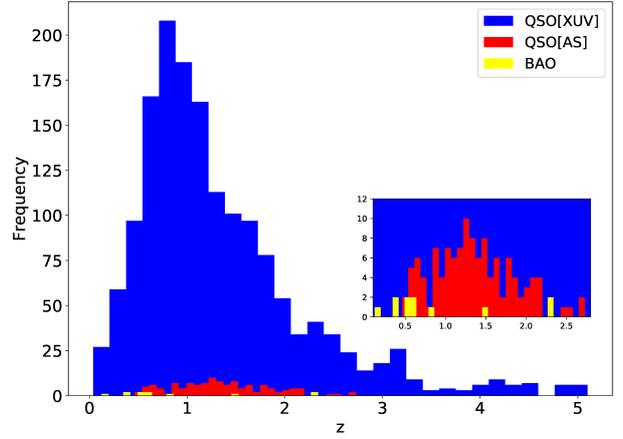}
\caption{The redshift distribution of the QSO and BAO measurements and its details (lower right).
Blue, red and yellow histograms stand for the redshift distribution
of QSO[XUV], QSO[AS], and BAO respectively.}
\end{figure}

\section{COSMOLOGICAL MODELS}

In this paper, we concentrate on four non-standard cosmological
models in a spatially flat universe, including the DGP model,
Generalized Chaplygin Gas (GCG) model, the MPC model, and the
power-law $f(T)$ model, based on Teleparallel Gravity.

\subsection{Dvali-Gabadadze-Porrati model}

Arising from the braneworld theory, the Dvali-Gabadadze-Porrati
model \citep{Dvali2000} modifies the gravity to reproduce the cosmic
acceleration without need to invoke DE. In this model, we are living
on a 4D membrane in a higher-dimensional spacetime. Moreover, the
gravity leaks out into the bulk at large scales, which will result
in the accelerated expansion of the Universe \citep{Li2013}. The
Friedman equation is modified as
\begin{equation}
H^{2}-\frac{H}{r_{c}}=\frac{8\pi G}{3}\rho _{m},
\end{equation}
where $r_{c}=1/[H_{0}(1-\Omega _{m})]$ represents the length scale
beyond which the leaking occurs. We can directly rewrite the above
equation and get the expansion rate
\begin{equation}
\frac{H(z)^{2}}{H_{0}^{2}}= \left( \sqrt{\Omega _{m}(1+z)^{3}+\Omega
_{rc}}+\sqrt{\Omega _{rc}} \right)^{2},
\end{equation}
where $H_{0}$ is the Hubble constant and $\Omega
_{rc}=1/(4r_{c}^{2}H_{0}^{2})$ is related to the cosmological scale.
Setting $z = 0$ in Eq. (2), the normalization condition can be
obtained
\begin{equation}
\Omega _{rc}=\frac{(1-\Omega _{m})^{2}}{4}.
\end{equation}
and there are two free parameters ${\hat p} = (\Omega _{m},H_{0})$
to be constrained.

\subsection{Generalized Chaplygin Gas model}

As one of the candidates for dark energy models, the Chaplygin gas
model, where dark energy and dark matter are unified through an
exotic equation of state, has been proposed to explain the cosmic
acceleration
\citet{Kamenshchik2001,Bento2002,Biesiada2005,Malekjani2011}. In
this model, the universe is filled with the so-called Chaplygin gas,
which is a perfect fluid characterized by the equation of state
$p=-A/\rho$. A more general case, in which
\begin{equation}
 p=-\frac{A}{\rho ^{\alpha }},
 \end{equation}
where $A$ is a positive constant and $\rho$ 
is the energy density of this fluid, is called the generalized
Chaplygin gas. The GCG model with $\alpha=0$ reduces to standard
$\Lambda$CDM model and with $\alpha=1$ reduces to the standard
Chaplygin gas (SCG) model. In the framework of FLRW metric, applying
Eq.~(4) and the conservation equation $d(\rho a^{3})=-pd(a^{3})$,
the energy density of GCG model is written as
\begin{equation}
\rho _{GCG}=\rho _{0GCG}\left [ A_{s}+(1-A_{s})a^{-3(1+\alpha )}
\right ]^{\frac{1}{1+\alpha}} ,
\end{equation}
where $a$ is the scale factor, $A_{s}=A/\rho _{0GCG}^{1+\alpha }$
and $\rho _{0GCG}$ is the present energy density of the GCG. Using
Eq. (4) and Eq. (5), one obtains the equation of state parameter of
GCG model
\begin{equation}
\omega_{GCG}=-\frac{A_{s}a^{3(1+\alpha
)}}{1-A_{s}+A_{s}a^{3(1+\alpha)}}.
\end{equation}
Eq. (6) shows clearly that GCG acts like dust matter
($\omega_{GCG}\rightarrow 0$) in the early time ($a\rightarrow 0$)
and behaves like a cosmological constant ($\omega_{GCG}\rightarrow
-1$) at late epoch ($a\rightarrow \infty$). The Friedman equation
for this model can be expressed as
\begin{equation}
\begin{split}
H(z)^{2}/H_{0}^{2}=&\Omega _{b}(1+z)^{3}+\\
&(1-\Omega _{b})\left [ A_{s}+(1-A_{s})(1+z)^{3(1+\alpha)} \right
]^{\frac{1}{1+\alpha }},
\end{split}
\end{equation}
where $\Omega _{b}$ is the present density parameter of the
baryonic matter. We adopt $100\Omega
_{b}h^{2}=2.166\pm0.015\pm0.011$ with $h=H_0/100$ as usual  and in
the uncertainty budget first term is associated with the deuterium
abundance measurement and the second one -- with the Big Bang
Nucleosynthesis (BBN) calculation used to get $\Omega _{b0}$
\citep{Cooke2018}. Since the parameter $A_{s}$ can be expressed by
the effective total matter density $\Omega _{m}$ and the $\alpha$
parameter
\begin{equation}
A_{s}=1-\left ( \frac{\Omega_m - \Omega_b}{1 - \Omega_b} \right
)^{1+\alpha },
\end{equation}
there are three free parameters ${\hat p} = (\Omega
_{m},\alpha,H_{0})$ in this model.

\subsection{Power-law f(T) model}

Lately, another kind of modified gravity theory-$f(T)$
\citep{Bengochea2009,Cai2016,Qi2017} proposed in the framework of
the Teleparallel Equivalent of General Relativity, has attracted a
lot of attention. In this scenario, the Weitzenböck connection with
torsion is used instead of torsionless Levi-Civita connection with
curvature used in General Relativity. The Lagrangian density is a
function $f(T)$ of the torsion scalar $T$, which is responsible for
the cosmic acceleration. In this framework, the Friedman equation
could be expressed as
\begin{equation}
\frac{H(z)^{2}}{H_{0}^{2}}=\Omega _{m}(1+z)^{3}+\Omega _{F}y(z,P),
\end{equation}
where $\Omega _{F}=1-\Omega _{m}$ and $y(z,{\hat p})$ can be written
as
\begin{equation}
y(z,{\hat p})=\frac{1}{T_{0}\Omega _{F}}(f-2Tf_{T}),
\end{equation}
with $T_{0}=-6H_{0}^{2}$, $f_{T}\equiv df/dT$, and ${\hat p}$
representing the parameters occurring in different forms of $f(T)$
theory. In this paper, we focus on the power-law $f(T)$ model with
the following form
\begin{equation}
f(T)=\alpha (-T)^{^{b}},
\end{equation}
where $\alpha$ and $b$ are two model parameters. The distortion
parameter $b$ quantifies the deviation from the $\Lambda$CDM model,
while the parameter $\alpha$ can be expressed through the Hubble
constant and density parameter $\Omega _{F0}$ by combining Eq. (9)
and Eq. (11) with the boundary condition $H(z=0)/H_{0}=1$:
\begin{equation}
\alpha=(6H_{0}^{2})^{1-b}\frac{\Omega _{F0}}{2b-1}.
\end{equation}
Now, Eq. (10) can be rewritten as
\begin{equation}
y(z,{\hat p})=E^{2b}(z,b).
\end{equation}
Here we consider the Taylor expansion up to second order for Eq.
(9), on $H(z,b)^{2}/H_{0}^{2}$ around $b=0$, to calculate the
Friedman equation (details can be found in \citet{Nesseris2013}).
Eventually, the free parameters in this $f(T)$ model are ${\hat p} =
(\Omega _{m},b,H_{0})$.

\subsection{Modified Polytropic Cardassian model}

In order to explain the accelerated cosmological expansion from a
different perspective, \citet{Freese2002} introduced the original
Cardassian model motivated by the braneworld theory, without DE
involved. In this model the Friedman equation is modified to
\begin{equation}
H^{2}=\frac{8\pi G\rho _{m}}{3}+B\rho _{m}^{n},
\end{equation}
where $\rho _{m}$ is the total matter density and the second term on
the right-hand side represents the Cardassian term. It is worth
noting that the universe is driven to accelerate by the Cardassian
term when the parameter $n$ satisfies $n<2/3$. Then, a simple
generalized case of the Cardassian model was proposed by
\citet{Gondolo2002,Wang2003}, where an additional exponent $q$ was
introduced. We can write the Friedman equation with this
generalization as
\begin{equation}
\frac{H(z)^{2}}{H_{0}^{2}}=\Omega _{m}(1+z)^{3}\times \left [
1+\left (\left (\frac{1}{\Omega _{m}}  \right )^{q}-1   \right
)(1+z)^{3q(n-1)}\right]^{1/q}.
\end{equation}
The MPC model, with the free parameters of ${\hat p} = (\Omega
_{m},n,q,H_{0})$ in this model, will reduce to $\Lambda$CDM model
when $q=1$ and $n=0$.

\section{METHODS}

In this section, we present the details of deriving observational
constraints on the cosmological models from QSOs and BAO
measurements.

\subsection{Quasars measurements}

Over the decades, a non-linear relation between the UV and X-ray
luminosities of quasars have been recognized and refined
\citep{Risaliti2015}. This relation can be expressed as
\begin{equation}
\log(L_X)=\gamma \log(L_{UV})+\beta,
\end{equation}
where $\log = \log_{10}$ and the slope -- $\gamma$ along with the
intercept -- $\beta$ are two free parameters, which should be
constrained by the measurements. Applying the flux-luminosity
relation of $F=L/{4\pi D_L^2}$, the UV and X-ray luminosities can be
replaced by the observed fluxes:
\begin{equation}
\log(F_X)=\gamma \log(F_{UV})+2(\gamma-1)\log
(D_L)+(\gamma-1)\log(4\pi)+\beta,
\end{equation}
where $F_{X}$ and $F_{UV}$ are the X-ray and UV fluxes,
respectively. Here $D_{L}$ is the luminosity distance, which
indicates such kind of QSO measurements can be used to calibrate
them as standard candles. Theoretically, $D_L$  is determined by the
redshift $z$ and cosmological parameters ${\hat p}$ in a specific
model:
\begin{equation}
D_{L}(z,{\hat p})=\frac{c(1+z)}{H_{0}}\int_{0}^{z}\frac{d
z'}{E(z')},
\end{equation}
where $E(z)\equiv H(z)/H_{0}$. In order to constrain cosmological
parameters ${\hat p}$ through the measurements of QSO X-ray and UV
fluxes, we compare the observed X-ray fluxes with the predicted
X-ray fluxes calculated with Eq. (17) at the same redshift. Then,
the best-fitted parameter values and respective uncertainties for
each cosmological model are determined by minimizing the $\chi ^{2}
= - 2\ln(LF)$ objective function, defined by the log-likelihood
\citep{Risaliti2015}:
    \begin{equation}
    \ln(LF)=-\frac{1}{2}\sum_{i=1}^{1598}\left [\frac{[\log(F_{X,i}^{obs})-\log(F_{X,i}^{th})]^{2}}{s_{i}^{2}}+ln(2\pi s_{i}^{2}) \right ],
    \end{equation}
    where $\ln=\log_{e}$, $s_{i}^{2}=\sigma_{i}^{2}+\delta^{2}$, and In  $\sigma_{i}$ is the measurement error on $F_{X,i}^{obs}$. In addition to the cosmological model parameters, three more free parameters are fitted: $\gamma$, $\beta$ representing the X-UV relation and $\delta$ representing the global intrinsic dispersion. Then, according to \citep{Khadka2020b}, for the purpose of model comparison we use the value of
     \begin{equation}
    \chi_{XUV, min}^{2}=-2\ln(LF)_{min}-\sum_{i=1}^{1598}\ln(2\pi (\sigma _{i,XUV}^{2}+\delta_{best\;fit} ^{2})).
    \end{equation}

In our analysis, another QSO data set comes from a new compiled
sample of 120 intermediate luminosity quasars \citep{Cao17a,Cao17b}
covering the redshift range $0.46  < z < 2.76$ with angular sizes
$\theta _{obs}(z)$, while the intrinsic length of this standard
ruler is calibrated to $l_m = 11.03 \pm 0.25$ pc through a new
cosmology-independent calibration technique \citep{Cao17b}. The
corresponding theoretical predictions for the angular sizes at
redshift $z$ can be expressed as
\begin{equation}
\theta_{th} (z)=\frac{l_{m}}{D_{A}(z)},
\end{equation}
where $D_{A}(z)$ is the angular diameter distance at redshift $z$
and
\begin{equation}
D_{A}(z)=\frac{D_{L}(z)}{(1+z)^{2}}.
\end{equation}
Then, one can derive model parameters by minimizing the $\chi ^{2}$
objective function:
\begin{equation}
\chi_{AS}^{2}(z;{\hat p})=\sum_{i=1}^{120}\frac{\left [ \theta
_{th}(z_{i};{\hat p})-\theta _{obs}(z_{i})\right ]^{2}}{\sigma
_{\theta }(z_{i})^{2}},
\end{equation}
where ${\hat p}$ denote free parameters in a specific cosmological
model and $\theta _{th}(z_{i};{\hat p})$ represents the theoretical
value of angular sizes at redshift $z_{i}$. Moreover, an additional
$10\%$ systematical uncertainty is added in the total uncertainty
$\sigma _{\theta }(z_{i})^{2}$ to account for the intrinsic spread
in the linear size \citep{Cao17b}. Therefore, in our analysis, the
total uncertainty is written as $\sigma _{\theta }(z_{i})^{2}=\sigma
_{\theta ,stat}(z_{i})^{2}+\sigma _{\theta ,sys}(z_{i})^{2}$, where
$\sigma _{\theta ,stat}(z_{i})^{2}$ is the statistical uncertainty
of $\theta _{obs}(z_{i})$ measurements.

\subsection{Baryon Acoustic Oscillations measurements}

For inclusion of the BAO measurements to the determination of
cosmological parameters, we follow the approach carried out in
\citet{Ryan2019}. It is well known that the BAO data, in particular
those listed in Table 1, are scaled by the size of the sound horizon
at the drag epoch $r_{s}$, which can be expressed as (details can be
found in \citet{Eisenstein1998})
\begin{equation}
r_{s}=\frac{2}{3k_{eq}}\sqrt{\frac{6}{R_{eq}}}\ln\left [
\frac{\sqrt{1+R_{d}}+\sqrt{R_{d}+R_{eq}}}{1+\sqrt{R_{eq}}} \right ],
\end{equation}
where $R_{d}$ and $R_{eq}$ are the values of the baryon to photon
density ratio
\begin{equation}
R=\frac{3\rho _{b}}{4\rho _{\gamma }},
\end{equation}
at the drag and matter-radiation equality redshifts $z_{d}$ and
$z_{eq}$, respectively, and $k_{eq}$ is the particle horizon
wavenumber at $z_{eq}$. The detailed expression of $z_{d}$,
$z_{eq}$, $k_{eq}$ and the baryon to photon density radio $R$ can be
found in \citet{Eisenstein1998}.

The BAO measurements listed in Table 1 involve the transverse
comoving distance (equal to the line of sight comoving distance if
$\Omega_{k0}=0$ )
\begin{equation}
D_{M}(z)= D_{c}(z) =
\frac{c}{H_{0}}\int_{0}^{z}\frac{d{z}'}{E({z}')}
\end{equation}
the expansion rate $H(z)$, angular diameter distance $D_A(z) =
\frac{D_M(z)}{1+z}$ and the volume-averaged angular diameter
distance
\begin{equation}
D_{V}(z)=\left [ \frac{cz}{H_{0}}\frac{D_{M}^{2}(z)}{E(z)} \right
]^{1/3}.
\end{equation}

For the measurements of the sound horizon ($r_{s}$) scaled by its
fiducial value, we use Eq. (24) to calculate both $r_{s}$ and
$r_{s,fid}$, following the approach applied in \citet{Ryan2019}. The
parameters of $(\Omega_{m},H_{0},\Omega _{b}h^{2})$ in the fiducial
cosmology are used as input to compute $r_{s,fid}$ where the BAO
measurements are reported. For the analysis that scales the BAO
measurements only by $r_{s}$, we turn to the fitting formula of
\citet{Eisenstein1998}, which is modified with a multiplicative
scaling factor of 147.60 Mpc$/r_{s,Planck}$ According to the
analysis of \citet{Ryan2019}, such modifications to the output of
the fitting formula may result in precise determinations of the size
of the sound horizon $r_{s}$ and $r_{s,fid}$. Let us note that the
baryon density $\Omega _{b}h^{2}$ is required to calculate the sound
horizon $r_{s}$ in Eq. (24). For the uncorrelated BAO measurements
listed in Table 1 (i.e. lines 7-9), the $\chi ^{2}$ objective
function can be written as
\begin{equation}
\chi_{BAO}^{2}({\hat p})=\sum_{i=1}^{3}\frac{\left [ A
_{th}(z_{i};{\hat p})-A _{obs}(z_{i})\right ]^{2}}{\sigma
(z_{i})^{2}},
\end{equation}
where $A _{th}$ and $A _{obs}$ are the predicted and measured
quantities of the BAO data listed in Table 1, and $\sigma (z_{i})$
stands for the relevant uncertainty of $A _{obs}$.

The BAO measurements listed in the first six lines and the last two
lines of Table 1 are correlated and consequently the $\chi^{2}$
objective function takes the form
\begin{equation}
\chi_{BAO}^{2}({\hat p})=\left [ A _{th}({\hat p})-A _{obs}\right
]^{T}C^{-1}\left [ A _{th}({\hat p})-A _{obs}\right ],
\end{equation}
where $C^{-1}$ denotes the inverse covariance matrix
\citep{Ryan2019}) for the BAO data taken from \citet{Alam2017},
while the covariance matrix is presented in \citet{Cao2020} for the
BAO data taken from \citet{de2019}.

\subsection{JOINT ANALYSIS}

We will perform the joint analysis of the above described data to
determine constraints on the parameters of a given model. In this
section we outline the underlying methodology. Using the $\chi ^{2}$
objective function defined above, one can write the likelihood
function as
\begin{equation}
\mathcal{L}({\hat p})=e^{- \frac{\chi ({\hat p})^{2}}{2}},
\end{equation}
where ${\hat p} $ is the set of model parameters under
consideration. Then, the likelihood function of the above combined
analysis is expressed as
\begin{equation}
\mathcal{L}=\mathcal{L}_{XUV}\mathcal{L}_{AS}\mathcal{L}_{BAO}.
\end{equation}
The likelihood analysis is performed using the Markov chain Monte
Carlo (MCMC) method, implemented in the \emph{emcee} package
\footnote{https://pypi.python.org/pypi/emcee} in Python 3.7
\citep{Foreman-Mackey2013}.

After constraining the parameters of each model, it is essential to
determine which model is most preferred by the observational
measurements and carry out a good comparison between the  different
models. Out of possible model selection techniques, we will use the
Akaike Information Criterion (AIC) \citep{Akaike1974}
    \begin{equation}
    AIC=\chi _{min}^{2}+2k,
    \end{equation}
as well as the Bayesian Information Criterion (BIC)
\citep{Schwarz1978}
    \begin{equation}
    BIC=\chi _{min}^{2}+klnN,
    \end{equation}
where $\chi _{min}^{2}=-2ln\mathcal{L}_{max}$, $k$ is the number of
free parameters in the model and $N$ represents the number of data
points. Moreover, the ratio of $\chi _{min}$ to the number of
degrees of freedom $dof=N-k$, is reported as an estimate of the
quality of the observational data set. The Akaike weights $\omega
_{i}(AIC)$ and Bayesian weights $\omega _{i}(BIC)$ are computed
through the normalized relative model likelihoods, which are
expressed as
\begin{equation}
\omega _{i}(IC)=\frac{exp\left \{ -\frac{1}{2}\Delta _{i}(IC) \right
\}}{\sum_{k=1}^{K}exp\left \{ -\frac{1}{2}\Delta _{k}(IC) \right
\}},
\end{equation}
where $\Delta _{i}(IC)$ is the difference of the value of given
information criterion IC (AIC or BIC) between the model $i$ and the
one which has the lowest IC and K denotes the total number of the
models considered. One can find the details of the rules for
estimating the AIC and BIC model selection in
\citet{Biesiada2007,Lu2008}.

We supplement the model comparison by calculating the Jensen-Shannon
divergence (JSD) \citep{Lin1991,Abbott2019} between the posterior
distributions of the common parameters  assessed with two different
cosmological models. The JSD is a symmetrized and smoothed measure
of the distance between two probability distributions $p(x)$ and
$q(x)$ defined as
\begin{equation}
D_{JS}(p\mid q)=\frac{1}{2}\left [ D_{KL}(p\mid s) +D_{KL}(q\mid
s)\right ],
\end{equation}
where $s=1/2(p+q)$ and $D_{KL}$ is the Kullback Leibler divergence
(KLD) between the distributions $p(x)$ and $q(x)$ expressed as
\begin{equation}
D_{KL}(p\mid q)=\int p(x) \log_{2}\left (\frac{p(x)}{q(x)}  \right
)dx,
\end{equation}
and a smaller value of the JSD indicates that the posteriors from
two models agree well \citep{Abbott2019}.

It should be pointed out that in order to compare models through the
JSD, we should use the posterior distributions of parameters ${\hat
p}$ which are the same in the models compared. Therefore, the matter
density $\Omega _{m}$ and the Hubble constant $H_{0}$ are the two
parameters of interest in our analysis. In addition, we will compare
the models described in Sec. 3 with $\Lambda$CDM model. Concerning
the posterior distributions of common free parameters in different
models, they can be obtained through the MCMC method, then we take
advantage of the dedicated Python 3.7 package
\footnote{scipy.spatial.distance.jensenshannon} to compute the JSD
between two one-dimensional (1D) probability distributions.


\begin{table*}
    \begin{center}
        \setlength{\tabcolsep}{1mm}{
        \begin{tabular}{|c|c|c|c|c|c|c|c|c|}
            \hline
            model & data & $\Omega _{m}$ & $\beta$ & $\gamma$ & $\delta$ & $H_{0}$ \\
            \hline
            $\Lambda$CDM & QSO[XUV]+QSO[AS] & $0.406_{-0.082}^{+0.108}$ & $7.321_{-0.314}^{+0.308}$ & $0.631_{-0.010}^{+0.010}$ & $0.231_{-0.004}^{+0.004}$ & $64.704_{-3.744}^{+3.347}$\\

            & BAO & $0.316_{-0.020}^{+0.022}$ & - & - & - & $68.074_{-1.380}^{+1.545}$\\

            & QSO[XUV]+QSO[AS]+BAO & $0.317_{-0.007}^{+0.007}$ & $7.152_{-0.260}^{+0.265}$ & $0.637_{-0.009}^{+0.009}$ & $0.231_{-0.004}^{+0.004}$ & $68.157_{-0.487}^{+0.496}$\\
            \hline
            model & data & $\Omega _{m}$ & $\beta$ & $\gamma$ & $\delta$ & $H_{0}$ \\
            \hline
            DGP & QSO[XUV]+QSO[AS] & $0.365_{-0.100}^{+0.129}$ & $7.358_{-0.309}^{+0.328}$ & $0.630_{-0.010}^{+0.010}$ & $0.231_{-0.004}^{+0.004}$ & $62.353_{-3.709}^{+3.877}$\\

            & BAO & $0.269_{-0.020}^{+0.022}$ & - & - & - & $59.560_{-1.004}^{+1.149}$\\

            & QSO[XUV]+QSO[AS]+BAO & $0.329_{-0.009}^{+0.009}$ & $7.299_{-0.271}^{+0.263}$ & $0.632_{-0.009}^{+0.009}$ & $0.231_{-0.004}^{+0.004}$ & $62.757_{-0.471}^{+0.488}$\\
            \hline
            model & data & $\Omega _{m}$ & $\alpha$ & $\beta$ & $\gamma$ & $\delta$ & $H_{0}$ \\
            \hline
            GCG & QSO[XUV]+QSO[AS] & $0.416_{-0.068}^{+0.088}$ & $2.360_{-1.793}^{+1.803}$ & $7.419_{-0.340}^{+0.326}$ & $0.628_{-0.011}^{+0.011}$ & $0.231_{-0.004}^{+0.004}$ & $69.254_{-4.970}^{+4.427}$\\

            & BAO & $0.299_{-0.039}^{+0.029}$ & $-0.227_{-0.246}^{+0.272}$ & - & - & - & $63.972_{-5.866}^{+5.266}$\\

            & QSO[XUV]+QSO[AS]+BAO & $0.319_{-0.009}^{+0.010}$ & $-0.067_{-0.147}^{+0.151}$ & $7.186_{-0.250}^{+0.259}$ & $0.636_{-0.009}^{+0.008}$ & $0.231_{-0.004}^{+0.005}$ & $67.496_{-1.904}^{+1.605}$\\
            \hline
            model & data & $\Omega _{m}$ & $b$ & $\beta$ & $\gamma$ & $\delta$ & $H_{0}$ \\
            \hline
            f(T) & QSO[XUV]+QSO[AS] & $0.409_{-0.090}^{+0.131}$ & $-0.193_{-0.509}^{+0.551}$ & $7.344_{-0.331}^{+0.313}$ & $0.631_{-0.010}^{+0.011}$ & $0.231_{-0.005}^{+0.005}$ & $63.829_{-4.839}^{+4.147}$\\

            & BAO & $0.303_{-0.029}^{+0.028}$ & $0.261_{-0.323}^{+0.252}$ & - & - & - & $64.154_{-4.858}^{+4.806}$\\

            & QSO[XUV]+QSO[AS]+BAO & $0.320_{-0.009}^{+0.010}$ & $0.084_{-0.166}^{+0.176}$ & $7.163_{-0.269}^{+0.280}$ & $0.637_{-0.009}^{+0.009}$ & $0.231_{-0.004}^{+0.005}$ & $67.507_{-1.987}^{+1.327}$\\
            \hline
            model & data & $\Omega _{m}$ & $q$ & $n$ & $\beta$ & $\gamma$ & $\delta$ & $H_{0}$ \\
            \hline
            MPC & QSO[XUV]+QSO[AS] & $0.410_{-0.062}^{+0.085}$ & $1.936_{-0.982}^{+2.923}$ & $-0.547_{-0.981}^{+0.564}$ & $7.497_{-0.374}^{+0.343}$ & $0.626_{-0.011}^{+0.012}$ & $0.231_{-0.004}^{+0.005}$ & $70.576_{-4.435}^{+6.475}$\\

            & BAO & $0.304_{-0.034}^{+0.030}$ & $0.914_{-0.322}^{+0.381}$ & $0.120_{-0.320}^{+0.192}$ & - & - & - & $63.800_{-5.620}^{+5.177}$\\

            & QSO[XUV]+QSO[AS]+BAO & $0.321_{-0.011}^{+0.011}$ & $0.944_{-0.183}^{+0.237}$ & $0.022_{-0.205}^{+0.165}$ & $7.176_{-0.253}^{+0.290}$ & $0.636_{-0.010}^{+0.008}$ & $0.231_{-0.005}^{+0.005}$ & $67.462_{-2.324}^{+1.663}$\\
            \hline
        \end{tabular}}
        \caption{Summary of the best-fit values with their 1$\sigma$
            uncertainties concerning the parameters of all considered
            models. The results are obtained from the combined data sets of QSO[XUV]+QSO[AS],
            BAO and QSO[XUV]+QSO[AS]+BAO.}
    \end{center}
\end{table*}

\begin{table*}
    \begin{center}
        \setlength{\tabcolsep}{1.5mm}{
        \begin{tabular}{|c|c|c|c|c|c|c|c|c|c|c|} 
            \hline
            data & model & AIC & $\Delta$AIC & $\omega_{i}(AIC)$ & BIC & $\Delta$BIC & $\omega_{i}(BIC)$ & $\chi_{min}^{2}/dof$ & $D_{JS}(\Omega_{m})$ & $D_{JS}(H_{0})$\\
            \hline
            QSO[XUV]+QSO[AS] & $\Lambda$CDM & 2217.95 & 1.25 & 0.257 & 2245.19 & 1.25 & 0.3419 & 1.289 & 0 & 0\\
            & DGP & 2216.70 & 0 & 0.479 & 2243.94 & 0 & 0.6376 & 1.288 & 0.233 & 0.273\\
            & GCG & 2218.58 & 1.88 & 0.188 & 2251.27 & 7.32 & 0.0164 & 1.289 & 0.199 & 0.447\\
            & f(T) & 2221.43 & 4.73 & 0.045 & 2254.12 & 10.18 & 0.0039 & 1.291 & 0.161 & 0.136\\
            & MPC & 2222.20 & 7.38 & 0.031 & 2260.34 & 16.40 & 0.0002 & 1.291 & 0.224 & 0.516\\
            \hline
            BAO & $\Lambda$CDM & 13.88 & 1.57 & 0.233 & 14.68 & 1.57 & 0.2454 & 1.098 & 0 & 0 \\
            & DGP & 12.32 & 0 & 0.511 & 13.11 & 0 & 0.5370 & 0.924 & 0.749 & 0.999\\
            & GCG & 15.95 & 3.63 & 0.083 & 17.14 & 4.03 & 0.0716 & 1.244 & 0.339 & 0.658\\
            & f(T) & 14.71 & 2.39 & 0.155 & 15.90 & 2.79 & 0.1332 & 1.090 & 0.270 & 0.640\\
            & MPC & 18.99 & 1.57 & 0.018 & 20.59 & 7.47 & 0.0128 & 1.571 & 0.255 & 0.663\\
            \hline
            QSO[XUV]+QSO[AS]+BAO & $\Lambda$CDM & 2227.20 & 0 & 0.780 &2254.47 & 0 & 0.9483 &1.286 & 0 & 0\\
            & DGP & 2233.63 & 6.43 & 0.031 &2260.90 & 6.43 & 0.0381 &1.289 & 0.566 & 0.999\\
            & GCG & 2230.45 & 3.25 &  0.153 &2263.18 & 8.71 & 0.0122  &1.288 & 0.198 & 0.586\\
            & f(T) & 2234.92 & 7.72 & 0.016 &2267.65 & 13.18 & 0.0013 &1.290 & 0.214 & 0.547\\
            & MPC & 2234.58 & 7.38 & 0.020 &2272.77 & 18.30 & 0.0001 &1.290 & 0.322 & 0.629\\
            \hline
        \end{tabular}}
        \caption{Information theoretic model comparison. Minimum values of
            AIC, BIC, their differences and weights are reported for the
            $\Lambda$CDM and each of the four cosmological models considered.
            Jensen-Shannon divergence $D_{JS}$ between $\Lambda$CDM and other
            cosmological models was calculated with respect to the matter
            density parameter $\Omega_{m}$ and the Hubble constant $H_{0}$.}
    \end{center}
\end{table*}

\section{RESULTS AND DISCUSSION}

In this section, we present the results for the four cosmological
models listed in Sec. 3, obtained using different combination of
data sets: QSO[XUV]+QSO[AS], BAO and QSO[XUV]+QSO[AS]+BAO. In order
to have a good comparison, the corresponding results for the
concordance $\Lambda$CDM model is also displayed. The 1D probability
distributions and 2D contours with $1\sigma$ and $2\sigma$
confidence levels, as well as the best-fit value with 1 $\sigma$
uncertainty for each model are shown in Figs. 2-6 and reported in
Table 2.

\begin{figure}
    \includegraphics[width=\columnwidth]{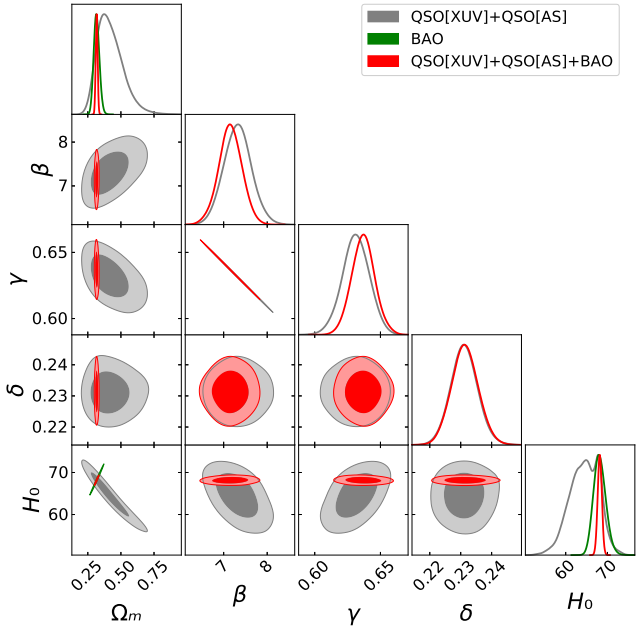}
    \caption{The 1D probability distributions and 2D contours with $1\sigma$
        and $2\sigma$ confidence levels for $\Lambda$CDM model obtained from
        QSO[XUV]+QSO[AS] (gray), BAO (green) and QSO[XUV]+QSO[AS]+BAO (red) data.}
\end{figure}

\subsection{Observational Constraints on Dvali-Gabadadze-Porrati model}

As can be seen from Fig. 3, the combined QSO measurements
(QSO[XUV]+QSO[AS]) do not provide stringent constraints on the
matter density parameter $\Omega_{m}$, which will be improved with
the combination of recent BAO observations. The best-fit value of
$\Omega _{m}$ given by QSO[XUV]+QSO[AS] is $\Omega
_{m}=0.365_{-0.100}^{+0.129}$ within 68.3\% confidence level, which
agrees well with the QSO[AS] data alone: $\Omega
_{m}=0.285_{-0.155}^{+0.255}$ (without systematics) \citep{Cao17b},
the recent Planck 2018 results: $\Omega _{m}=0.315\pm 0.007$
\citep{Planck2018} and SNe Ia+BAO+CMB+observational Hubble parameter
(OHD): $\Omega _{m}=0.305\pm 0.015$ \citep{Shi2012}. However, it is
worthwhile to mention that the matter density parameter $\Omega
_{m}$ obtained by QSO tends to be higher than that from other
cosmological probes, as was remarked in the previous works of
\citet{Risaliti2019,Khadka2020b}. This suggests that the composition
of the universe characterized by cosmological parameters can be
comprehended differently through high-redshift quasars. For the BAO
data, the best-fit matter density parameter is
$\Omega_{m}=0.269_{-0.020}^{+0.022}$, which is significantly lower
than that from the modified gravity theories considered in this
paper. Interestingly, the estimated values of $\Omega_{m}$ are in
agreement with the standard ones reported by other astrophysical
probes, such as $\Omega_{m}=0.277_{-0.017}^{+0.017}$ given by the
linear growth factors combined with CMB+BAO+SNe+GRB observations
\citep{Xia2009}, $\Omega_{m}=0.235_{-0.074}^{+0.125}$ given by
galaxy clusters combined with SNe+GRBs+CMB+BAO+OHD observations
\citep{Liang2011}, and $\Omega_{m}=0.243_{-0.074}^{+0.077}$ given by
strong gravitational lensing systems \citep{Ma2019}. For comparison,
the fitting results from the combined QSO[XUV]+QSO[AS]+BAO data sets
are also shown in Fig. 3, with the the matter density parameter of
$\Omega_{m}=0.329_{-0.009}^{+0.009}$. The use of BAO data to
constrain cosmological models seems to be complementary to the QSO
distance measurements, considering the constrained results
especially on $\Omega _{m}$ and $H _{0}$.

\begin{figure}
    \includegraphics[width=\columnwidth]{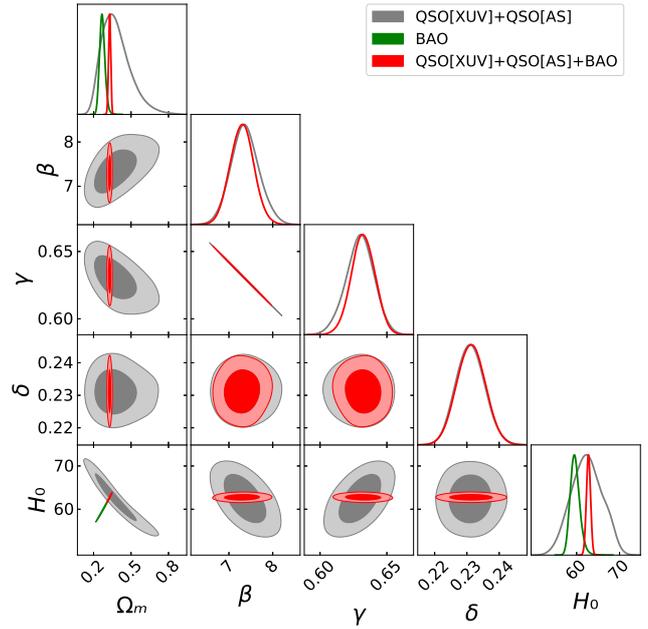}
    \caption{The 1D probability distributions and 2D contours with
        $1\sigma$ and $2\sigma$ confidence levels for DGP model obtained
        from QSO[XUV]+QSO[AS] (gray), BAO (green) and QSO[XUV]+QSO[AS]+BAO (red) data.}
\end{figure}

\subsection{Observational Constraints on Generalized Chaplygin Gas model}

Fig. 4 and Table 2 present the results of the best-fitted parameters
for the GCG model. One can see a deviation between the constraints
of ($\Omega _{m}$, $\alpha$, $H _{0}$) coming from the three
combined data sets, which are still consistent with each other
within $2\sigma$ confidence level. On the one hand, the combined
data sets QSO[XUV]+QSO[AS] can not tightly constrain the model
parameters ($\Omega _{m}$,$\alpha$), especially for parameter
$\alpha$ whose best-fit value is $\alpha=2.360_{-1.793}^{+1.803}$
and $\Omega _{m}$ is much larger than the value implied by other
measurements. In the framework of GCG, considering the fact that the
parameter $\alpha$ quantifies the deviation from the $\Lambda$CDM
model and the SCG model, $\Lambda$CDM is not consistent with GCG at
$1\sigma$ confidence level, while SCG is more favored by the
QSO[XUV]+QSO[AS] data. However, in the case of BAO and
QSO[XUV]+QSO[AS]+BAO data, $\Lambda$CDM is still favored within
$1\sigma$, with $\alpha=-0.227_{-0.246}^{+0.272}$ and
$\alpha=-0.067_{-0.147}^{+0.151}$ respectively. The combined data
set of QSO[XUV]+QSO[AS]+BAO provides more stringent constraints on
the matter density parameter ($\Omega _{m}=0.319_{-0.009}^{+0.010}$)
and the Hubble constant ($H_{0}=67.496_{-1.904}^{+1.605} \Mpc$). For
comparison, the results obtained from the joint light-curve analysis
(JLA) compilation of SNe Ia, CMB, BAO, and 30 OHD data simulated
over redshift range $2 \leqslant z  \leqslant 5$ gave $\Omega
_{m}=0.345_{-0.006}^{+0.006}$ and $\alpha=-0.047_{-0.026}^{+0.027}$
\citep{Liu2019}, which prefers a higher value of $\Omega _{m}$ than
our results and does not include $\Lambda$CDM within $1\sigma$
range. It is interesting to note that \citet{Liu2019} also obtained
$\alpha=-0.040_{-0.065}^{+0.060}$ without adding the simulated data,
which still includes the $\Lambda$CDM model at $1\sigma$ confidence
level and is slightly different from the results obtained by adding
the simulated higher redshift. This may indicate that the data
within the ``redshift desert'' ($2 \leqslant z  \leqslant 5$) can
provide a valuable supplement to other astrophysical observations in
the framework of GCG model.

\begin{figure}
    \includegraphics[width=\columnwidth]{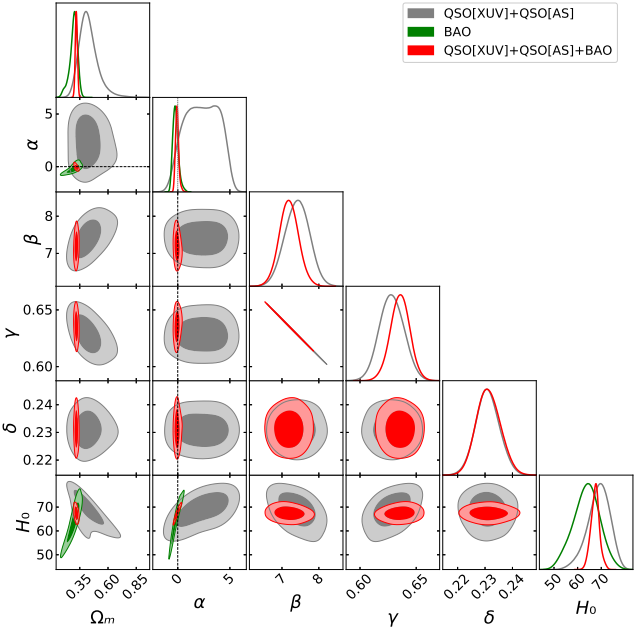}
    \caption{The 1D probability distributions and 2D contours with
        $1\sigma$ and $2\sigma$ confidence levels for GCG model obtained
        from QSO[XUV]+QSO[AS] (gray), BAO (green) and QSO[XUV]+QSO[AS]+BAO (red) data.
        The black dashed line represents the $\Lambda$CDM model
        corresponding to $\alpha=0$.}
\end{figure}

\subsection{Observational Constraints on Power-law F(T) model}

In the case of $f(T)$ theory based on $f(T)=\alpha (-T)^{^{b}}$
ansatz, the results are presented in Fig. 5 and can be seen in Table
2. It can be clearly seen from the comparison plots, there is a
consistency between QSO[XUV]+QSO[AS], BAO and QSO[XUV]+QSO[AS]+BAO.
However, the QSO[XUV]+QSO[AS] data generate a higher matter density
parameter $\Omega _{m}=0.409_{-0.090}^{+0.131}$ compared with other
probes. As for the parameter $b$ which captures the deviation of the
$f(T)$ model from the $\Lambda$CDM model, the best-fit value is
$b=-0.193_{-0.509}^{+0.551}$ and the $\Lambda$CDM model ($b=0$) is
still included within $1\sigma$ range. Such conclusion could also be
carefully derived in the case of BAO and QSO[XUV]+QSO[AS]+BAO
measurements. For comparison, our results are similar to the results
obtained with QSO(AS)+SNe Ia+BAO+CMB data sets ($\Omega
_{m}=0.317\pm 0.010$, $b=0.057_{-0.065}^{+0.091}$) \citep{Qi2017}
and SNe Ia+BAO+CMB+dynamical growth data ($\Omega _{m}=0.272\pm
0.008$, $b=-0.017\pm 0.083$) \citep{Nesseris2013}, where the value
of $\Omega _{m}$ is in tension with our results within $1\sigma$.
Moreover, in the framework of the power-law $f(T)$ model, the
parameter $b$ obtained from QSO[XUV]+QSO[AS] and BAO alone seems to
deviate from zero more according to the above mentioned results,
which suggests that there are still some possibility that
$\Lambda$CDM may not be the best cosmological model preferred by
current observations with larger redshift range. With the combined
data sets QSO[XUV]+QSO[AS]+BAO, we also get stringent constraints on
the model parameters $\Omega _{m}=0.320_{-0.010}^{+0.009}$,
$b=0.084_{-0.166}^{+0.176}$ and $H_{0}=67.507_{-1.987}^{+1.327}
\Mpc$, where $\Lambda$CDM is included within $1\sigma$. It is worth
noting that this slight deviation from the $\Lambda$CDM is also in
agreement with similar results in the literature, obtained from
QSO[AS]+BAO+CMB ($b=0.080\pm 0.077$) \citep{Qi2017} and OHD+SN
Ia+BAO+CMB ($b=0.05128_{-0.019}^{+0.025}$) \citep{Nunes2016}.

\begin{figure}
\includegraphics[width=\columnwidth]{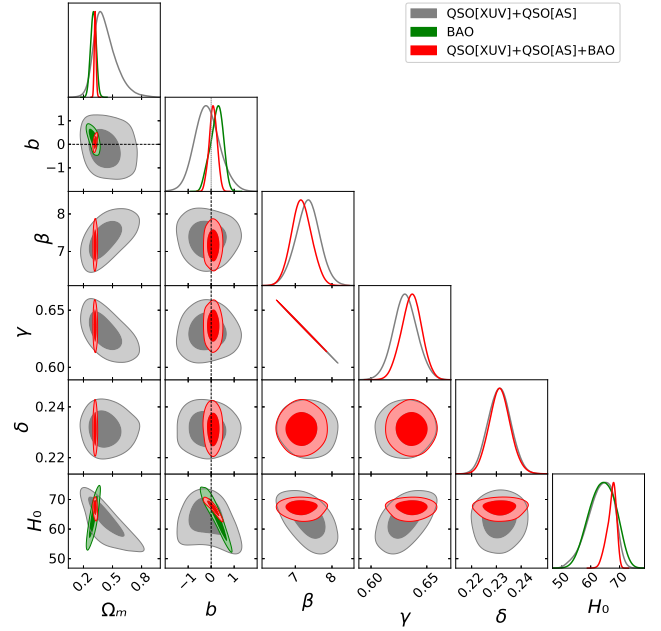}
\caption{The 1D probability distributions and 2D contours with
$1\sigma$ and $2\sigma$ confidence levels for $f(T)$ model obtained
from QSO[XUV]+QSO[AS] (gray), BAO (green) and QSO[XUV]+QSO[AS]+BAO
(red) data. The black dashed line represents the $\Lambda$CDM model
corresponding to $b=0$.}
\end{figure}

\subsection{Observational Constraints on Modified Polytropic Cardassian model}

All values of the estimated cosmic parameters in the MPC model are
displayed in Table 2 and illustrated in Fig. 6. For the
QSO[XUV]+QSO[AS], BAO and QSO[XUV]+QSO[AS]+BAO data, the best-fit
values of the parameters ($q$ and $n$) are in good agreement with
each other within within $2\sigma$. Meanwhile, $\Lambda$CDM is still
favored by the current QSO and BAO measurements within $1\sigma$
confidence level in the MPC model. Apparently, however, the combined
QSO data supports higher matter density parameter $\Omega
_{m}=0.410_{-0.062}^{+0.085}$ than that from the other two data sets
within 68.3\% confidence level. The constraints on the parameters
$q$ and $n$ from the QSO[XUV]+QSO[AS] data are very weak:
$q=1.936_{-0.982}^{+2.923}$, $n=-0.547_{-0.981}^{+0.564}$, but it
seems that the central values of $q$ and $n$ deviate more from 1 and
0, respectively in comparison to the results obtained with BAO and
QSO[XUV]+QSO[AS]+BAO. This may suggest, that the quasar data
(especially QSO[XUV]) at higher redshift may have some possibility
of favoring the modifications to the Friedmann equations in the MPC
model. Several authors have tested the MPC model with different
measurements. For instance, the SNe Ia+BAO+CMB+OHD data sets gave
$q=0.897_{-0.468}^{+0.152}$, $n=-0.648_{-1.106}^{+0.856}$
\citep{Shi2012}, which are in good accordance with our results
obtained of the BAO measurements and QSO[XUV]+QSO[AS]+BAO. In
addition, our limits are similar to $q=3.29\pm 3.30$, $n=0.26\pm
0.13$ shown in \citet{Magana2015} using BAO data, and in tension
with the results obtained from strong lensing measurements
\citep{Magana2015} in Abell1689: $q=5.2\pm 2.25$, $n=0.41\pm 0.25$.
With the SNLS3 SN Ia sample+CMB+BAO+OHD data sets, \citet{Li2012}
got the constraints $q=1.098_{-0.465}^{+1.015}$ and
$n=0.014_{-0.946}^{+0.364}$, which is consistent with our limits.
Note that, also with the QSO[XUV]+QSO[AS]+BAO data sets the best-fit
values of $q$ and $n$ parameters deviate from 1 and 0 respectively,
which implies the possibility of the modifications to the Friedmann
equations.

From our constraints on the matter density parameter $\Omega _{m}$
in different non-standard cosmological models, one thing is quite
clear, which is that the combined QSO data containing large number
of measurements at high redshifts ($2\leqslant z  \leqslant 5$) do
favor $\Omega _{m}$ lying in the range from
$0.365_{-0.100}^{+0.129}$ to $0.416_{-0.068}^{+0.088}$. This is
considerably higher than the constraints from other probes (such as
BAO measurements). Actually, such results on $\Omega _{m}$ have been
noted in the previous works through different approaches. For
instance, \citet{Khadka2020b} obtained $\Omega _{m} \sim 0.5-0.6$ in
four different cosmological models with large number of QSO[XUV],
while \citet{Qi2017} derived $\Omega _{m}=0.319\pm 0.011$ and
$\Omega _{m}=0.329\pm 0.011$ in different $f(T)$ theories with
QSO[AS]+BAO+CMB data sets. Other studies of different dark energy
models (based on Pade approximation parameterizations) revealed the
similar conclusions with Pantheon+GRB+QSO: the matter density
parameter lies in the range from $\Omega
_{m}=0.384_{-0.022}^{+0.033}$ to $\Omega
_{m}=0.391_{-0.026}^{+0.038}$ \citep{Rezaei2020}. Meanwhile, some
recent studies
\citep{Risaliti2019,Lusso2019,Rezaei2020,Benetti2019,Yang2020,Demianski2020,Li2021}
focused on exploring the deviation between high redshift
measurements and the standard cosmological model. Despite of the
$\Omega_m$ inconsistency obtained from the measurements with
different redshift coverage, it is still under controversy whether
this is an indication of a new physics or an unknown systematic
effect of the high-redshift observations. Therefore, besides
developing new high quality and independent cosmological probes, it
would be more interesting to figure out why the standard
cosmological parameters are fitted to different values with high and
low redshift observations. The latter indicates that one could go
beyond $\Lambda$CDM model to properly describe our universe
\citep{Ding15,Zheng16}.

As for the constraints on the Hubble constant $H_{0}$ shown in Table
2, one can see that $H_{0}$ lies in the range from
$H_{0}=62.353_{-3.709}^{+3.877}$ to
$H_{0}=70.576_{-4.435}^{+6.475}\Mpc$ for the combined QSO data, from
$H_{0}=59.560_{-1.004}^{+1.149}$ to
$H_{0}=68.074_{-1.380}^{+1.545}\Mpc$ for BAO, and from
$H_{0}=62.757_{-0.471}^{+0.488}$ to
$H_{0}=68.157_{-0.487}^{+0.496}\Mpc$ for QSO[XUV]+QSO[AS]+BAO.
Apparently, almost all the results for $H_{0}$ are lower than $70
\Mpc$, except for the MPC model assessed with QSO[XUV]+QSO[AS] which
however has large uncertainties.

\begin{figure}
    \includegraphics[width=\columnwidth]{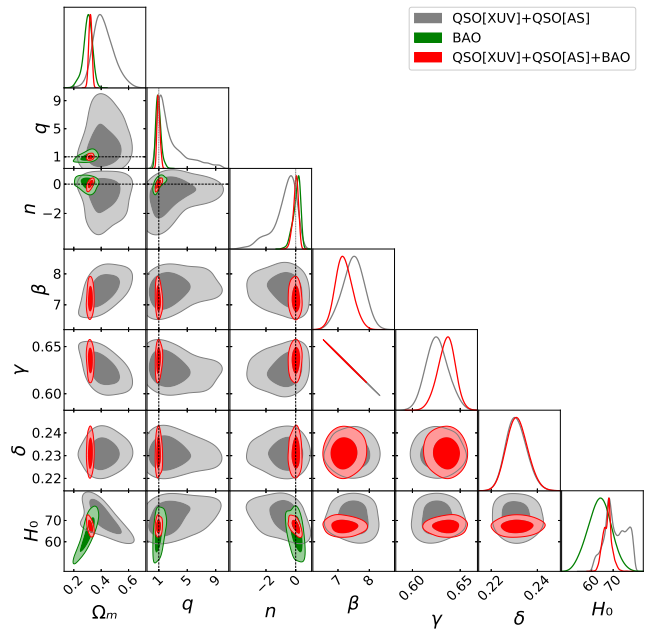}
    \caption{The 1D probability distributions and 2D contours with
        $1\sigma$ and $2\sigma$ confidence levels for MPC model, obtained
        from QSO[XUV]+QSO[AS] (gray), BAO (green) and QSO[XUV]+QSO[AS]+BAO (red) data.
        The black dashed line indicates the $\Lambda$CDM model corresponding
        to $q=1$ and $n=0$.}
\end{figure}

\begin{figure*}
    \begin{center}
        \subfloat[Combined QSO]{%
            \includegraphics[width=0.5\linewidth]{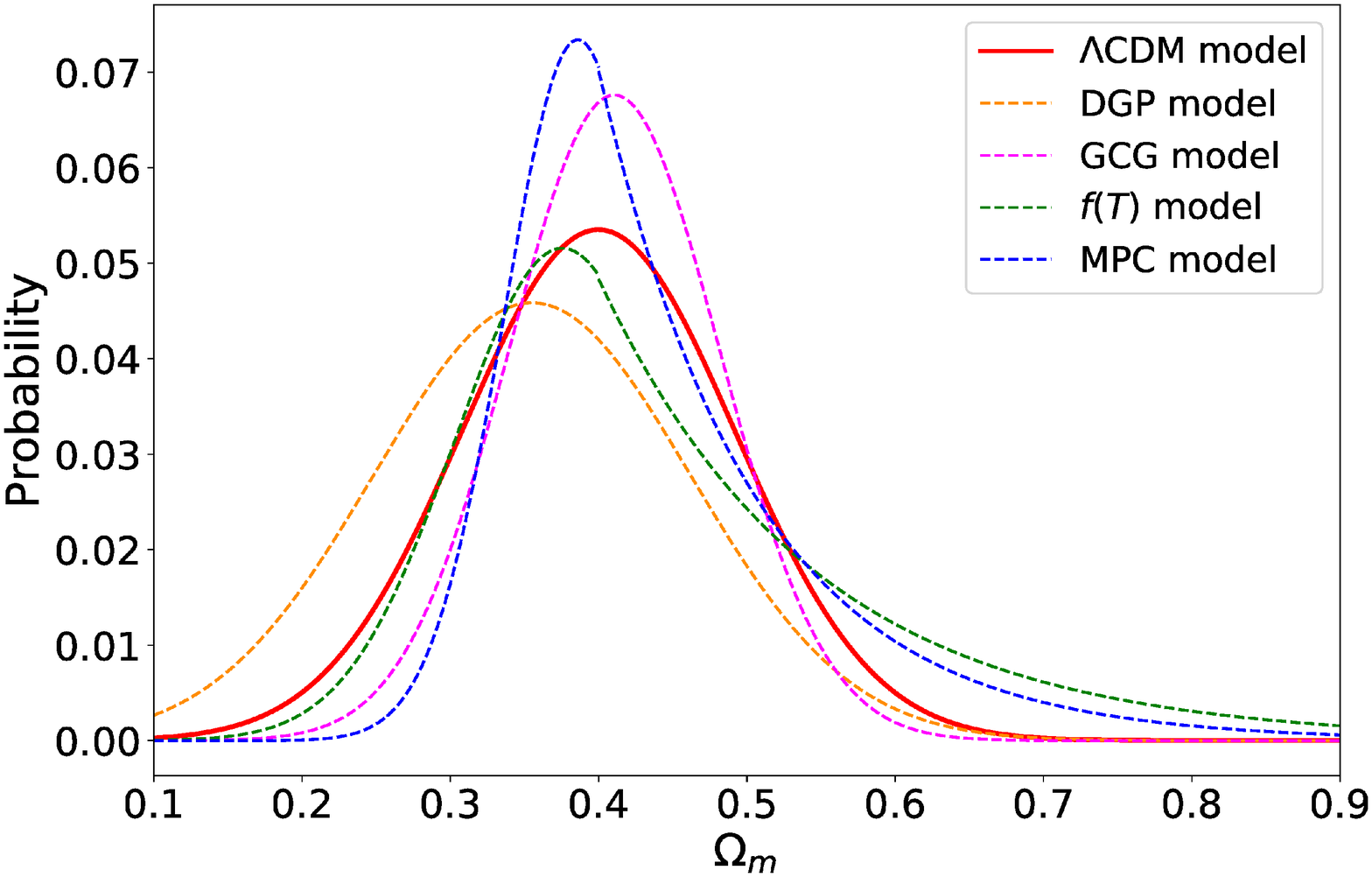}}\hspace{-5mm}
        \subfloat[BAO]{%
            \includegraphics[width=0.5\linewidth]{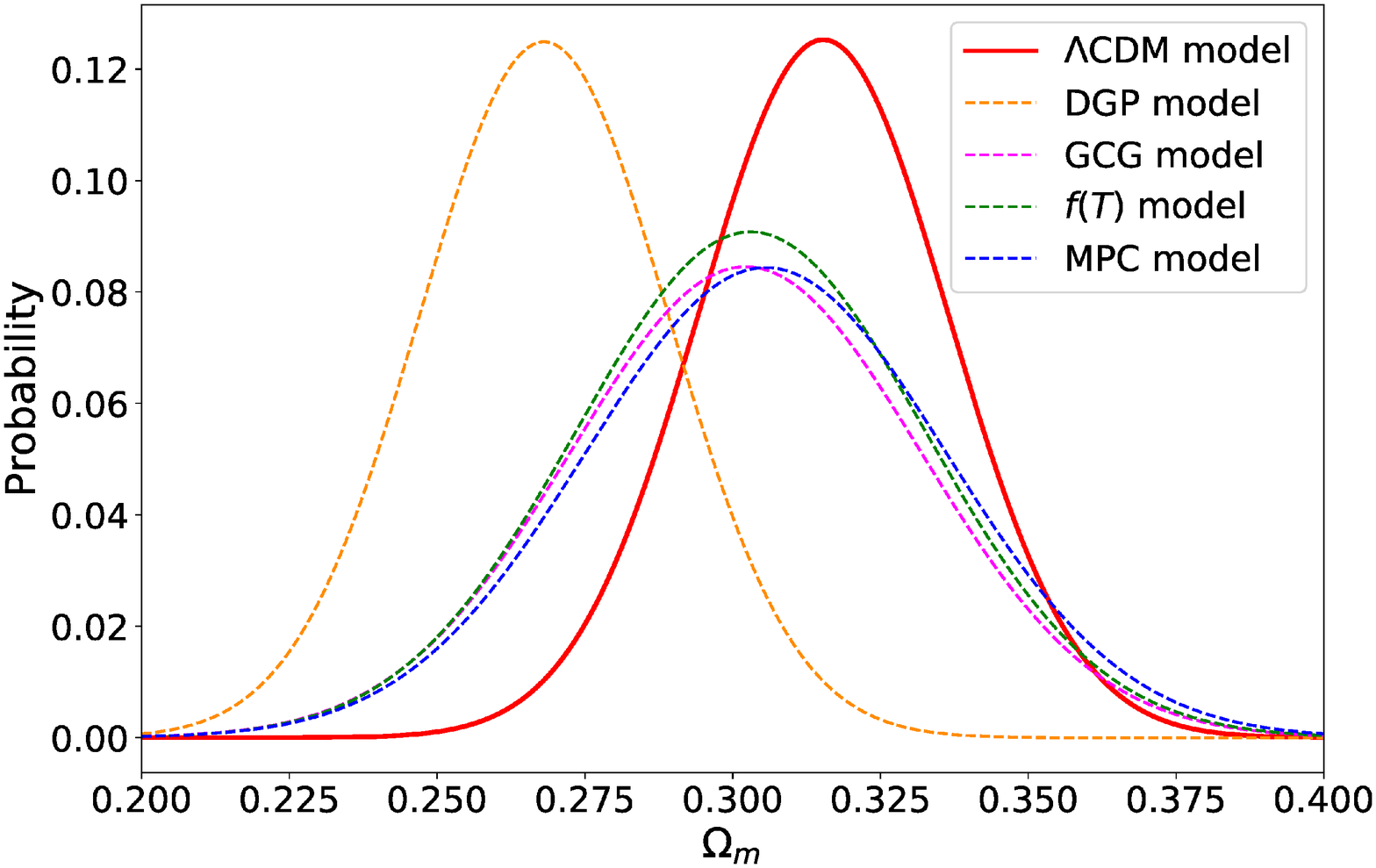}}\hspace{-5mm}
        \subfloat[Combined QSO and BAO]{%
            \includegraphics[width=0.5\linewidth]{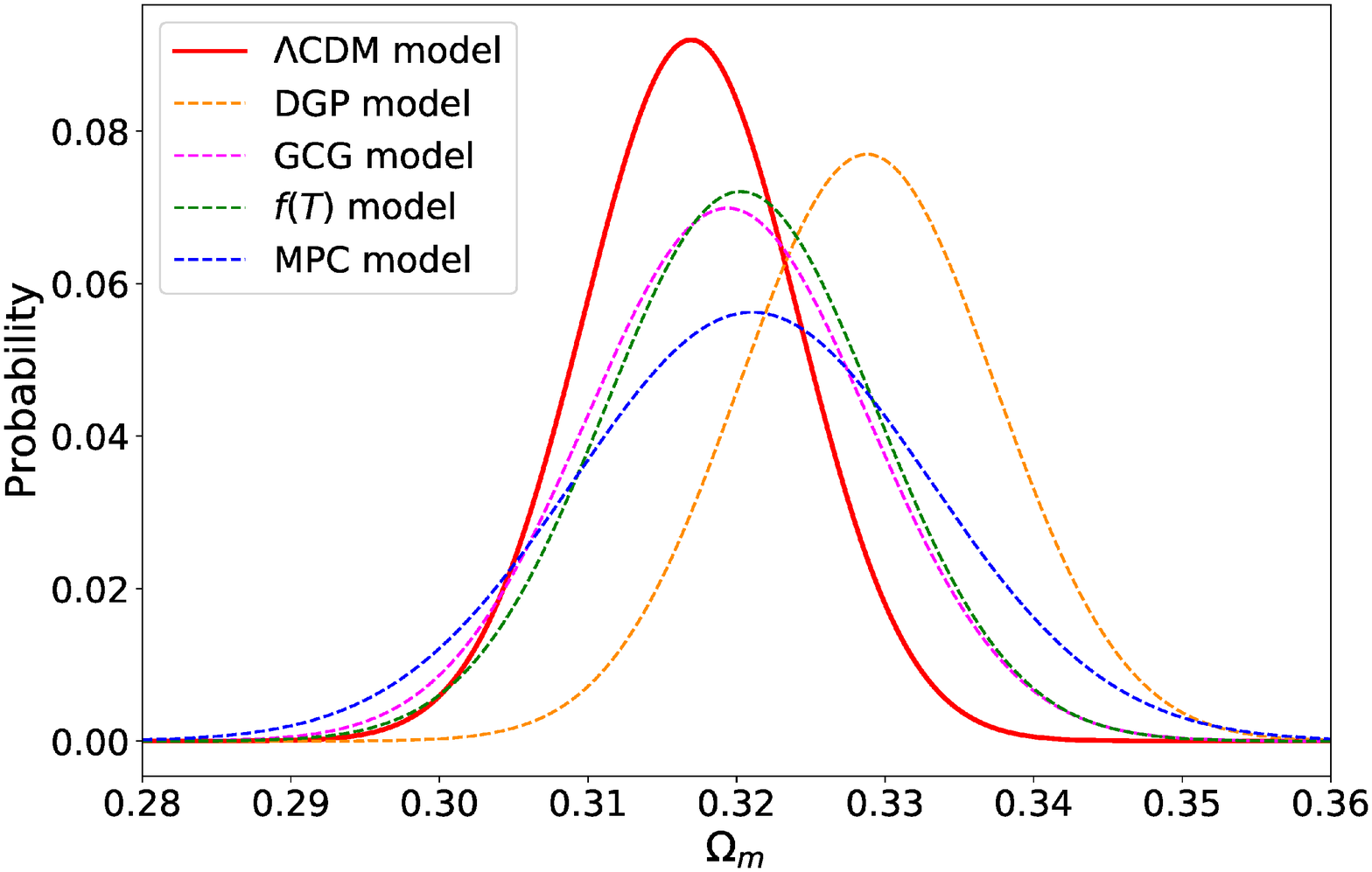}}
    \end{center}
\caption{The posterior distributions of $\Omega_{m}$ for
$\Lambda$CDM, DGP, GCG, $f(T)$, and MPC model, with the combined
QSO, BAO and combined QSO+BAO data.}
\end{figure*}

\subsection{MODEL COMPARISON}

In this section we compare the models and discuss how strongly are
they supported by the observational data sets. In Table 3 one can
find the summary of the information theoretical model selection
criteria applied to different models from QSO[XUV]+QSO[AS], BAO and
QSO[XUV]+QSO[AS]+BAO data sets. It can be seen that $\Lambda$CDM is
still the best model for the combined data QSO[XUV]+QSO[AS]+BAO,
under the assessment of AIC and BIC. Although the quasar sample
(QSO[XUV]+QSO[AS]) and the BAO data tend to prefer the DGP model in
term of AIC and BIC, they also share the same preference for
$\Lambda$CDM, compared with other theories of modified gravity.

It is important to keep in mind that model selection provides a
quantitative information on the strength of evidence (or the degree
of support) rather than just selecting only one model
\citep{Lu2008}. Table 3 informs us that AIC applied to the
QSO[XUV]+QSO[AS] data set does not effectively discriminate
$\Lambda$CDM and GCG models -- both of them receive the similar
support, while the evidence against $f(T)$ and MPC model is very
strong. For the QSO[XUV]+QSO[AS]+BAO data, we find that the DGP,
$f(T)$ and MPC model are clearly disfavored by the data, as they are
unable to provide a good fit. The BIC diversifies the evidence
between the models. Out of all the candidate models, it is obvious
that models with more free parameters (GCG, $f(T)$ and MPC) are less
favored by the current quasar observations (QSO[XUV]+QSO[AS]). Among
these four modified gravity models, the evidence against MPC is very
noticeable for all kinds of data sets, which demonstrates the MPC
model is seriously punished by the BIC.

Traditional information criteria (AIC or BIC) do not provide much
insight into the agreement between $\Lambda$CDM and the other four
models. Therefore, we also calculated the JSD (see Sec. 4.4) in
order to assess which models are consistent with $\Lambda$CDM in
light of the observational data. As already mentioned one should
have common parameters in all compared models and we used in this
role the matter density parameter and Hubble constant. Figs. 7-8
show the posterior distribution of $\Omega_{m}$ and $H_{0}$ obtained
with QSO[XUV]+QSO[AS], BAO and QSO[XUV]+QSO[AS]+BAO data sets for
all models considered. For QSO[XUV]+QSO[AS] data, the posterior
distributions of $\Omega_{m}$ and $H_{0}$ in $f(T)$ models agree
more with that of $\Lambda$CDM in terms of the value of JSD. As for
the BAO measurements, the value of JSD concerning $\Omega_{m}$ shows
the MPC model agree more with the $\Lambda$CDM, but concerning
$H_{0}$, all four non-standard models give large distance from the
$\Lambda$CDM, where the $f(T)$ model is still the closest to it. In
the case of QSO[XUV]+QSO[AS]+BAO data sets, the DGP and MPC model
are much more distant from the $\Lambda$CDM for the posterior
distributions of $\Omega_{m}$, while GCG and $f(T)$ model is closer
to it, which is similar to the cases for the posterior distributions
of $H_{0}$.

\begin{figure*}
    \begin{center}
        \subfloat[Combined QSO]{%
            \includegraphics[width=0.5\linewidth]{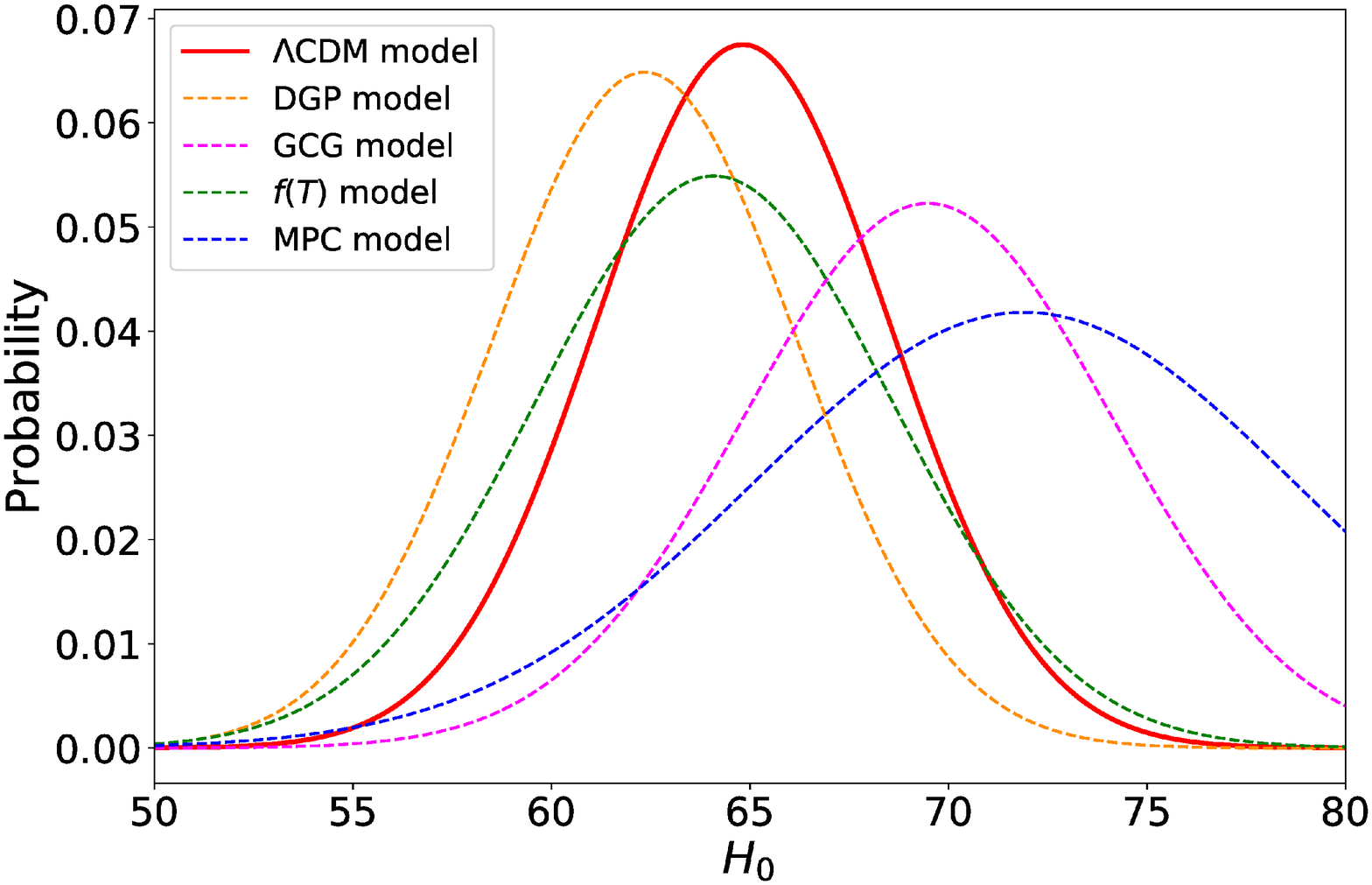}}\hspace{-5mm}
        \subfloat[BAO]{%
            \includegraphics[width=0.5\linewidth]{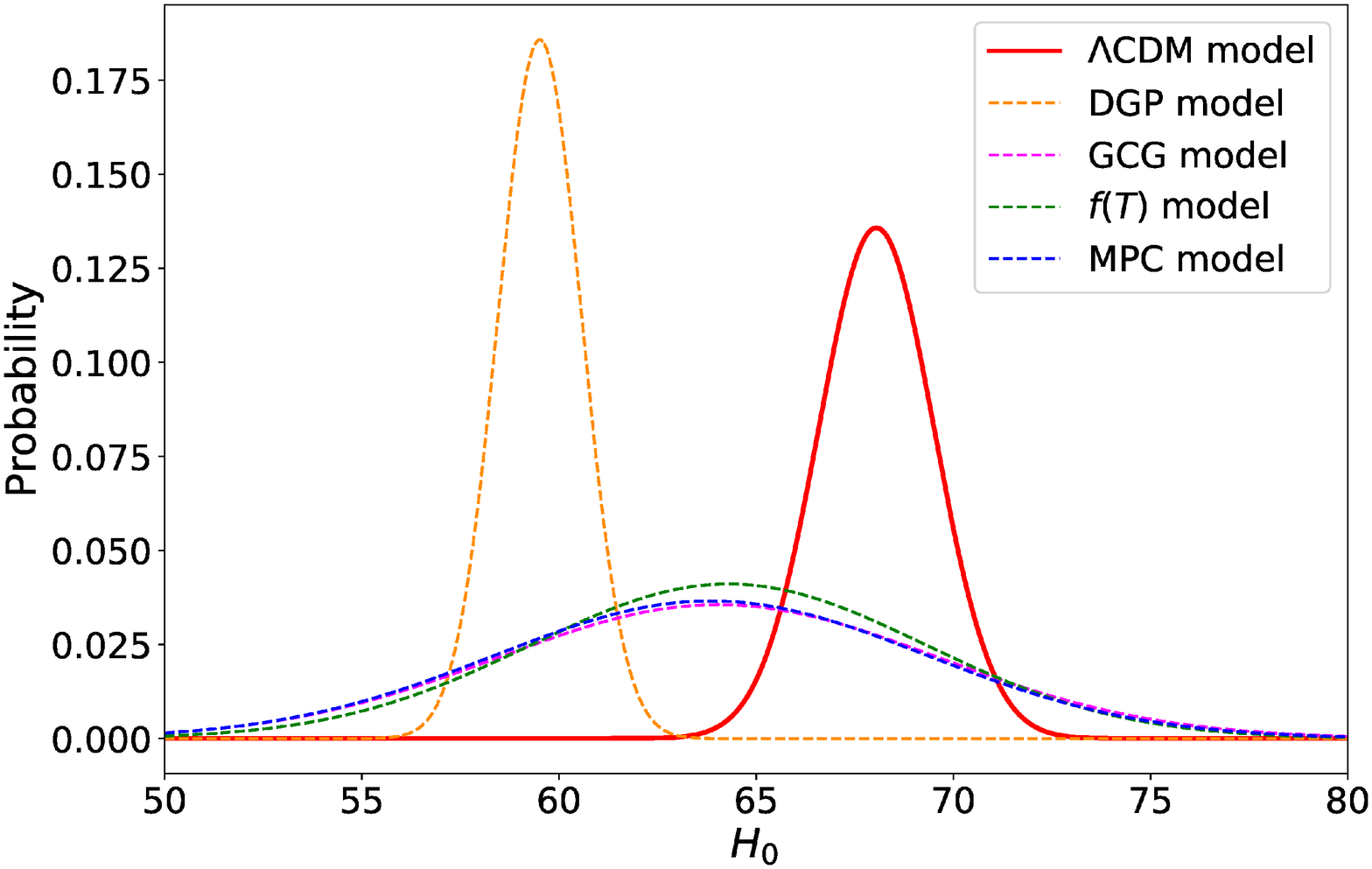}}\hspace{-5mm}
        \subfloat[Combined QSO and BAO]{%
            \includegraphics[width=0.5\linewidth]{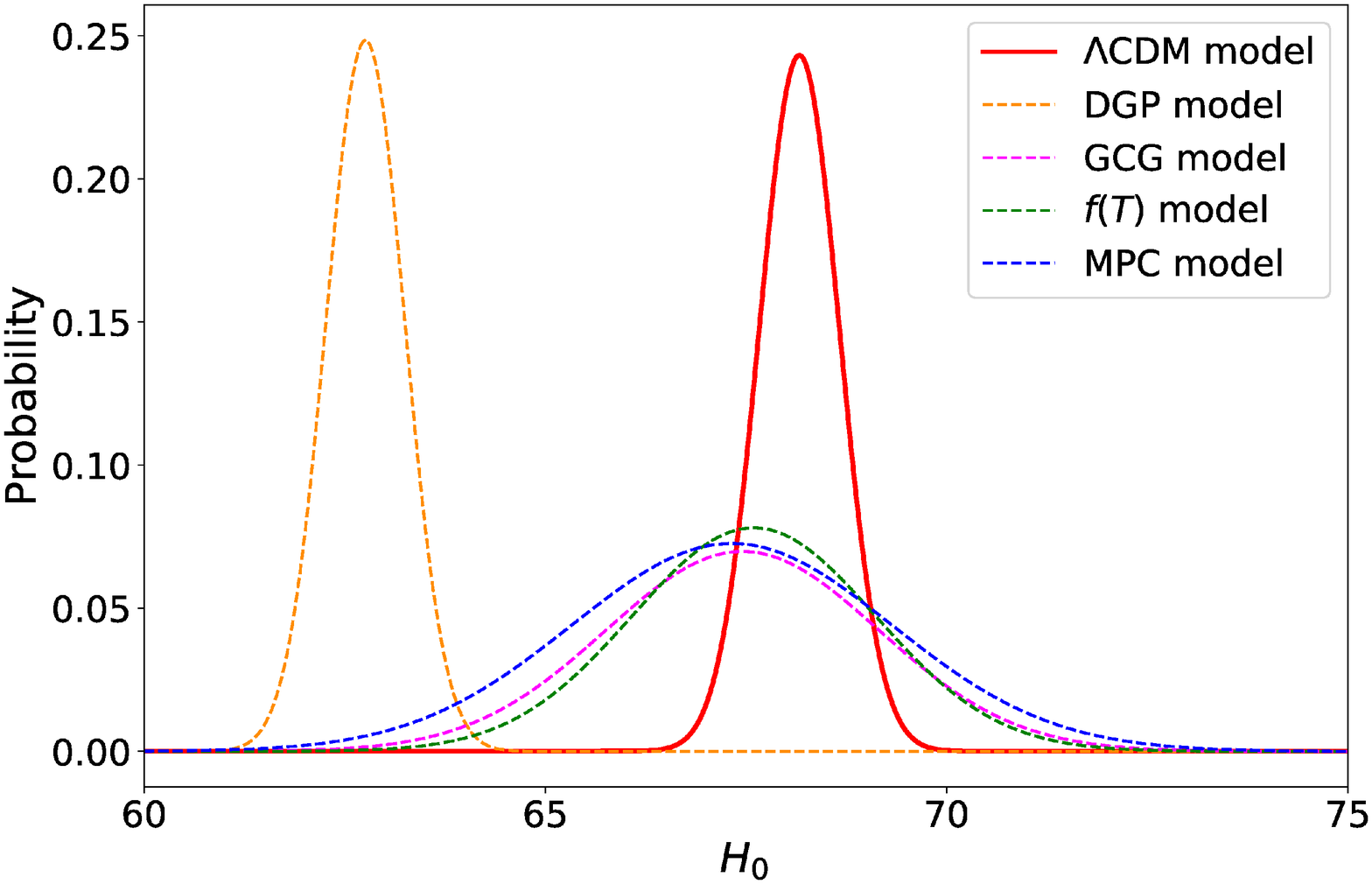}}
    \end{center}
\caption{The posterior distributions of  $H_{0}$ for $\Lambda$CDM,
DGP, GCG, $f(T)$, and MPC model, with the combined QSO, BAO and
combined QSO+BAO data.}
\end{figure*}

\section{CONCLUSIONS}

The modified gravity could provide interesting approaches to explain
the cosmic acceleration, without involving dark energy. In this
paper, we have evaluated the power of multiple measurements of
quasars covering sufficiently wide redshift range, on constraining
some popular modified gravity theories including DGP, GCG, the
power-law $f(T)$ model and MPC model, under the assumption of the
spatial flatness of the Universe. As for the observational data, the
newest large sample of QSO X-ray and UV flux measurements
\citep{Risaliti2019} were used as standard candles and provided a
good opportunity to test models at the "redshift desert" ($2
\leqslant z \leqslant 5$) that is not yet widely available through
other observations. In addition, a popular compilation of 120
angular size measurements of compact structure in radio quasars
versus redshift data from very-long-baseline interferometry (VLBI)
over the redshift range $0.46<z<2.76$ \citep{Cao17b} was used as
standard rulers to test these models in conjunction with the 1598
QSO X-Ray and UV flux measurements. Meanwhile, with the aim to
tighten the constraint from the combined QSO data sets and test the
consistency with other observations, 11 recent BAO measurements in
the redshift range $0.122\leq z\leq 2.34$ \citep{Cao2020} were also
taken into account in this work. Here we summarize our main
conclusions in more detail:

\begin{itemize}

\item Our results show that calibrating parameters $\beta$ and $\gamma$
from the non-linear $L_{X}-L_{UV}$ relation, as well as the global
intrinsic dispersion $\delta$ are almost independent of the
cosmological model, which is similar to the results from
\citet{Khadka2020b}. This supports the evidence that these selected
quasars can be regarded as standard candles.

\item In all four non-standard cosmological models, the results show that
the combined QSO data alone are not able to provide tight
constraints on model parameters, which is mainly related to the
large dispersion ($\delta =0.23$) of the $L_{X}-L_{UV}$ relation
obtained from the 1598 QSO X-Ray and UV flux measurements. On the
other hand, the combined quasar data constraints are mostly coherent
with the joint analysis including BAO measurements. The value of
matter density parameter $\Omega_{m}$ implied by the combined QSO
data is noticeably larger than that derived from other measurements,
which is likely caused by the discrepancy between the QSO X-Ray and
UV flux data and the $\Omega_{m}=0.3$ flat $\Lambda$CDM
\citep{Risaliti2019,Khadka2020b}. It is quite possible that quasar
data at high redshifts can shed new light on the model of our
universe. Moreover, in this paper, we tested different alternative
models. Most of them include the concordance $\Lambda$CDM model as a
special case corresponding to certain values of their parameters,
such as the parameter $b$ in the power-law $f(T)$ model. For the
$f(T)$ and MPC model, $\Lambda$CDM turned out to be compatible with
them at $1\sigma$ confidence level, while the GCG model is generally
inconsistent with the cosmological constant case within $1\sigma$.
Furthermore, after including BAO in the joint analysis, the best-fit
value of these parameters and their $1\sigma$ confidence levels show
less deviation from $\Lambda$CDM, which suggests that BAO
measurements favor $\Lambda$CDM significantly.

\item According to the AIC and BIC, the concordance $\Lambda$CDM model is still
the best cosmological model in light of the combined QSO and BAO
data, while the MPC model has considerably less support as the best
one. Although the quasar sample (QSO[XUV]+QSO[AS]) and the BAO data
tend to prefer the DGP model in term of AIC and BIC, they also share
the same preference for $\Lambda$CDM, compared with other theories
of modified gravity. Therefore, non-standard models with more free
parameters are less favored by the available observations, which is
the most unambiguous result of the current dataset. In order to
compare the agreement between the $\Lambda$CDM model and other four
models, Jensen-Shannon divergence (JSD) was applied in this paper.
We found that for the combined QSO data, the posterior distribution
of $\Omega_{m}$ and $H_{0}$ from $f(T)$ were in a better agreement
with $\Lambda$CDM. For BAO measurements, MPC and $f(T)$ models are
closer to $\Lambda$CDM according to the values of JSD from the
posterior distribution of $\Omega_{m}$, while all four models are
distant from $\Lambda$CDM in the case of $H_{0}$, especially the DGP
model. With QSO+BAO data, the results are similar to that from BAO
measurements, but the posterior distribution of $\Omega_{m}$ from
GCG and $f(T)$ model are in better agreement with $\Lambda$CDM.

\end{itemize}

\section*{acknowledgments}

This work was supported by the National Natural Science Foundation
of China under Grant Nos. 12021003, 11690023, 11633001 and
11920101003, the National Key R\&D Program of China (Grant No.
2017YFA0402600), the Beijing Talents Fund of Organization Department
of Beijing Municipal Committee of the CPC, the Strategic Priority
Research Program of the Chinese Academy of Sciences (Grant No.
XDB23000000), the Interdiscipline Research Funds of Beijing Normal
University, and the Opening Project of Key Laboratory of
Computational Astrophysics, National Astronomical Observatories,
Chinese Academy of Sciences. M.B. was supported by the Foreign
Talent Introducing Project and Special Fund Support of Foreign
Knowledge Introducing Project in China. He was supported by the Key
Foreign Expert Program for the Central Universities No. X2018002.

\section*{Data Availability Statements}

The data underlying this article will be shared on reasonable
request to the corresponding author.

{}

\end{document}